\documentclass[prl,twocolumn,amsmath,amssymb,showpacs,showkeys,superscriptaddress]{revtex4}
\usepackage{graphicx}
\usepackage{dcolumn}
\usepackage{hyperref}
\newcommand{\be}{\begin{equation}}
\newcommand{\ee}{\end{equation}}
\newcommand{\ba}{\begin{eqnarray}}
\newcommand{\ea}{\end{eqnarray}}

\begin{document}

\title{Phase diagram of the Gaussian-core model}

\author{Santi Prestipino}
\altaffiliation{Corresponding Author} \email{\tt
Santi.Prestipino@unime.it} \affiliation{Universit\`a degli Studi di
Messina,\\Dipartimento di Fisica, Contrada Papardo, 98166 Messina,
Italy}
\author{Franz Saija}
\email{\tt saija@me.cnr.it}
\affiliation{Istituto per i Processi Chimico-Fisici del CNR,
Sezione di Messina,\\Via La Farina 237, 98123 Messina, Italy}
\author{Paolo V. Giaquinta}
\email{\tt Paolo.Giaquinta@unime.it}
\affiliation{Universit\`a degli Studi di Messina,\\Dipartimento di
Fisica, Contrada Papardo, 98166 Messina, Italy}

\date{\today}

\begin{abstract}
We trace with unprecedented numerical accuracy the phase diagram of
the Gaussian-core model, a classical system of point particles
interacting via a Gaussian-shaped, purely repulsive potential. This
model, which provides a reliable qualitative description of the
thermal behavior of interpenetrable globular polymers, is known to
exhibit a polymorphic FCC-BCC transition at low densities and
reentrant melting at high densities. Extensive Monte Carlo
simulations, carried out in conjunction with accurate calculations
of the solid free energies, lead to a thermodynamic scenario that is
partially modified with respect to previous knowledge. In
particular, we find that: i) the fluid-BCC-FCC triple-point
temperature is about one third of the maximum freezing temperature;
ii) upon isothermal compression, the model exhibits a
fluid-BCC-FCC-BCC-fluid sequence of phases in a narrow range of
temperatures just above the triple point. We discuss these results
in relation to the behavior of star-polymer solutions and of other
softly repulsive systems.
\end{abstract}

\pacs{05.20.Jj, 61.20.Ja, 64.10.+h, 64.70.Kb}

\keywords{Gaussian-core model; Solid-liquid phase transitions;
Solid-solid phase transitions; Frenkel-Ladd method; Self-avoiding
polymers}

\maketitle

It is common knowledge that
crystallization is induced by the strong Pauli repulsion between
inner-shell electrons, causing the {\em effective} interatomic
potential to blow up at short distances. However, the existence of a
thermodynamically stable solid phase does not necessarily require a
singular repulsion for vanishing interatomic separations. As a
matter of fact, a finite square barrier, equal to a positive
constant $\epsilon$ for distances smaller than a given diameter
$\sigma$ being zero otherwise, is an example of a bounded repulsion
that supports a stable solid at all
temperatures.~\cite{Likos1,Schmidt} In this respect, a
Gaussian-shaped potential \be v(r)=\epsilon\exp(-r^2/\sigma^2)\,,
\label{eq1} \ee is a more realistic finite-strength repulsion.
The so-called Gaussian-core model (GCM) was originally introduced by
Stillinger.~\cite{Stillinger1} Such a potential, despite the fact that
it is finite even at full overlap between the particles, is nonetheless
perfectly admissible as an effective potential. Actually, it is used
to represent the entropic repulsion between (the centers of mass of)
self-avoiding polymer coils dispersed in a good
solvent.~\cite{Lang,Likos2,Louis} Two distinctive features of the
GCM, which are absent in the simpler penetrable-sphere system, are:
i) the existence of a maximum freezing temperature, $T_{\rm max}$;
ii) the occurrence, below $T_{\rm max}$, of reentrant melting into a
dense fluid phase. Stillinger noted in his original paper that, in
the limit of vanishing temperature and density, the GCM particles
practically behave as hard spheres with increasingly large diameter.
In this limit, the fluid freezes into a face-centered-cubic (FCC)
structure at a temperature $T_{\rm f}(\rho)$ that vanishes with the
number density $\rho$ (from now on, temperature and density will be
given in reduced units, $T^*=k_{\rm B}T/\epsilon$, where $k_{\rm B}$
is Boltzmann's constant, and $\rho^*=\rho\sigma^3$). Indeed, a
straightforward calculation of the total energy of different cubic
crystal structures shows that the FCC structure is favored, at zero
temperature, only for reduced densities lower than $\pi^{-3/2}\simeq
0.1796$. Beyond this threshold, a body-centered-cubic (BCC) solid
takes over. However, upon compression, any regular arrangement of
particles is eventually destined to collapse for any $T>0$
(reentrant melting).~\cite{Stillinger2}

A comprehensive study of the phase diagram of the GCM was recently
carried out by Lang and coworkers.~\cite{Lang} These Authors
employed an approximate integral-equation theory to describe the
disordered phase and a variationally-adjusted harmonic interaction
for the crystalline phases. The resulting phase diagram (see Fig.\,9
of Ref.\,\cite{Lang}) accounts for the existence of a fluid phase
and two solid phases. More specifically, the solid was found to be
thermodynamically stable for temperatures lower than $T_{\rm
max}^*=0.0102$, a temperature at which the freezing line attains its
maximum value for $\rho_{\rm max}^*=0.2292$. The phase diagram
displays a fluid-BCC-FCC triple point that was estimated to fall at
$T_{\rm tr}^*=0.00875$. The FCC phase turns out to be stable below
$T_{\rm tr}^*$ in the low-density region (for $\rho^*$ less than
$\approx 0.17$), whereas the BCC structure prevails for
larger temperatures and densities.

A parallel numerical study of the phase diagram of star-polymer
solutions in a good solvent was carried out under a different
assumption for the effective pair potential, modeled with
an ultrasoft logarithmic repulsion within a diameter $\sigma$ and
with a Yukawa potential outside the core.~\cite{Watzlawek} In
this case, Monte Carlo (MC) simulations and free-energy calculations
lead to a phase diagram that shows a rather complicate interplay
between various cubic phases. It is desirable to have a similarly
fully-fledged analysis also for the GCM.

To this purpose, we performed standard Metropolis MC simulations of
the GCM, keeping the number of particles $N$, the volume $V$, and
the temperature $T$ constant. We used the particle-insertion
method~\cite{Widom} and the Frenkel-Ladd technique~\cite{Frenkel,Polson}
to calculate the ``exact'' free energies of the dilute fluid and of the
solid phases, respectively.
By this means, we discovered that the phase diagram of the GCM is
more elaborate than previously reported,~\cite{Lang} showing
elements of similarity with the phase diagram of star-polymer
solutions.

%
%
\begin{table}
\caption{\label{tab1} Excess Helmholtz free energy per particle $f_{\rm ex}$,
in units of $k_{\rm B}T$, calculated for some FCC ($N=1372$) and BCC
($N=1458$) solid states of the GCM. For $T^*=0.003$ and $0.006$, the
tabulated values refer to systems with $864$ and $1024$ particles,
respectively. For each state and solid structure, we also display
(within square brackets) the value of the reduced elastic constant
$c^*=c\,\sigma^2/\epsilon$ that intervenes in the Frenkel-Ladd
calculation: for the given $c$, the mean square displacement of the
Einstein crystal approximately matches the mean square deviation of
a GCM particle from its reference crystal site. For a number of
selected $\rho^*$ and $T^*$, we verified that the quantity $\beta
f_{\rm ex}(N)+\ln N/N$, with $\beta=(k_{\rm B}T)^{-1}$, scales
linearly with $N^{-1}$ for large $N$, in agreement with a conjecture
formulated in \cite{Polson}.}
\begin{tabular*}{\columnwidth}[c]{@{\extracolsep{\fill}}|r|r|r|r|}
\hline
$\rho^*$ & $T^*$ & $\beta f_{\rm ex}^{\rm (FCC)}\,\,\,\,\,\,\,\,\,\,\,\,\,\,\,$ & $\beta f_{\rm ex}^{\rm (BCC)}\,\,\,\,\,\,\,\,\,\,\,\,\,\,\,$\\
\hline\hline
$0.30$ & $0.0020$ & 195.703(2)\,\,\,\,$[0.30]$ & 195.312(1)\,\,\,\,$[0.45]$ \\
\hline
$0.24$ & $0.0030$ & 86.251(2)\,\,\,\,$[0.35]$ & 86.057(1)\,\,\,\,$[0.39]$ \\
\hline
$0.24$ & $0.0033$ & 78.994(1)\,\,\,\,$[0.34]$ & 78.814(1)\,\,\,\,$[0.38]$ \\
\hline
$0.24$ & $0.0035$ & 74.835(1)\,\,\,\,$[0.33]$ & 74.666(1)\,\,\,\,$[0.38]$ \\
\hline
$0.24$ & $0.0037$ & 71.122(2)\,\,\,\,$[0.33]$ & 70.961(1)\,\,\,\,$[0.38]$ \\
\hline
$0.24$ & $0.0038$ & 69.411(2)\,\,\,\,$[0.32]$ & 69.254(1)\,\,\,\,$[0.37]$ \\
\hline
$0.30$ & $0.0040$ & 101.074(2)\,\,\,\,$[0.29]$ & 100.894(1)\,\,\,\,$[0.42]$ \\
\hline
$0.24$ & $0.0060$ & 46.025(2)\,\,\,\,$[0.29]$ & 45.929(1)\,\,\,\,$[0.35]$ \\
\hline
$0.24$ & $0.0080$ & 35.781(2)\,\,\,\,$[0.24]$ & 35.710(1)\,\,\,\,$[0.31]$ \\
\hline
\end{tabular*}
\end{table}

Our samples typically consisted of 1372 particles for the fluid and
the FCC solid, and of 1458 particles for the BCC solid.
Occasionally, we considered smaller as well as larger sizes, so as
to check whether our conclusions were possibly undermined by a
significant finite-size dependence. We payed much care to a safe
estimate of statistical errors. This is actually an important issue
whenever different crystalline structures so closely compete, as in
the present case, for thermodynamic stability. We computed
free-energy differences between two equilibrium states of the
system belonging to the same phase through standard thermodynamic
integration. This method allows one to obtain the properties of the
model for any state, provided that the absolute free energy has
been autonomously computed in at least one reference state per
phase. Table 1 gives the excess Helmholtz free energy for some FCC and BCC
states of the model. Though we did not systematically check the
relative stability of other crystalline structures, we verified
that, for temperatures close to the triple point, the
hexagonal-close-packed (HCP) solid is slightly less favored than
the FCC solid, while the simple-cubic solid is not mechanically
stable.

\begin{figure}
\includegraphics[width=8cm,angle=0]{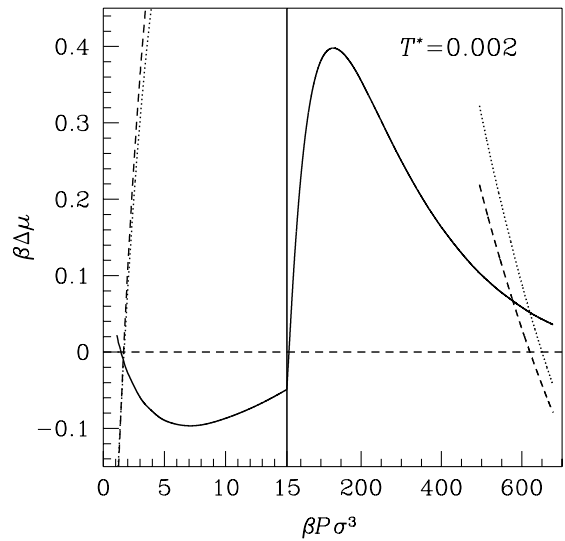}
\caption{\label{fig1} Difference between the chemical potentials of
pairs of GCM phases plotted as a function of the pressure along the
isotherm $T^*=0.002$: $\beta\Delta\mu_{\rm fluid,\,FCC}$ (dashed
line), $\beta\Delta\mu_{\rm fluid,\,BCC}$ (dotted line), and
$\beta\Delta\mu_{\rm FCC,\,BCC}$ (continuous line). Upon increasing $P$
(or $\rho$), the fluid transforms into a FCC solid; then, a FCC-BCC
transition takes place until the BCC melts into a fluid phase again.
The lines are spline interpolants of the data points. A zoom on the
low-pressure region shows the non-monotonic behavior of
$\beta\Delta\mu_{\rm FCC,\,BCC}$, a feature that is ultimately
responsible, at higher temperatures, for the reentrance of the
BCC phase (see also Fig.\,\ref{fig3}).}
\end{figure}

The thermodynamically stable phase, for given temperature and
pressure, is the one with lowest chemical potential $\mu(T,P)$.
Figure\,\ref{fig1} shows the difference between the chemical
potentials of competing phases plotted as a function of $P$ at fixed
temperature. The sequence of phase transitions undergone by the GCM
at $T^*=0.002$, with increasing pressures, is fluid-FCC-BCC-fluid:
the disordered phase is actually seen to reenter the phase diagram
at high density. As is also apparent from Fig.\,\ref{fig1}, the
$\mu$ gap between the FCC and BCC phases is anything but monotonous
when plotted as a function of the pressure. Note that, at low
pressures, the BCC phase is about to become stable as the fluid
freezes into a FCC structure. In fact, this eventually occurs at
higher temperatures, in a way not documented before for the GCM.

%
%

%
%
\begin{figure}
\includegraphics[width=8cm,angle=0]{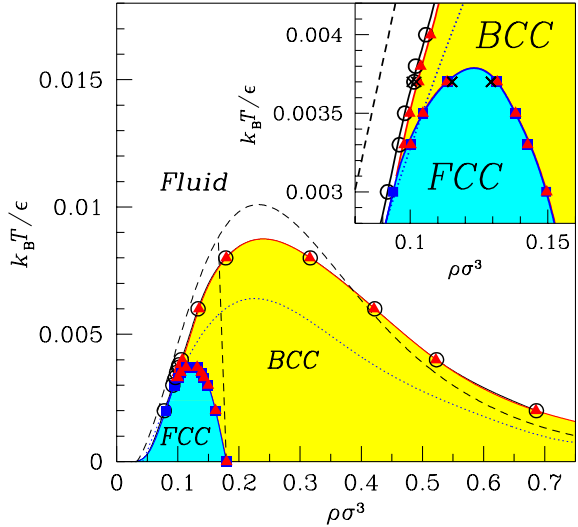}
\caption{\label{fig2} Phase diagram of the GCM in the $(\rho,T)$
plane, with a zoom on the triple-point region (inset). Transition
densities for each phase are shown for various temperatures: fluid
($N=1372$, black open circles), FCC ($N=1372$, blue solid squares),
and BCC ($N=1458$, red solid triangles). Black crosses refer to
simulations of larger systems ($N=2048$, fluid and FCC; $N=2000$,
BCC) at $T^*=0.0037$. Continuous lines drawn through the data points
are a guide for the eye. We also plotted the freezing and FCC-BCC
coexistence loci calculated in \cite{Lang} (black dashed lines), and
the ordering threshold predicted by a one-phase entropy-based
criterion (blue dotted line).~\cite{Giaquinta}}
\end{figure}

Upon increasing the temperature, the shallow valley in $\mu_{\rm
FCC}-\mu_{\rm BCC}$ moves gradually upwards until a narrow range of
pressure appears (for $T^*\gtrsim 0.0030$) where a stable BCC phase
slips in between the fluid and the FCC solid phases. This is
possible because the corresponding increase of the freezing pressure
with temperature is not large enough to suppress the reentrant BCC
phase. For $T^*>0.0038$, the FCC phase ceases to be stable and a
more regular behavior sets in, similar to that predicted by Lang and
coworkers.~\cite{Lang}

All in all, the phase diagram represented in Fig.\,\ref{fig2}
emerges. If compared with Fig.\,9 of \cite{Lang}, two differences
stand out: a definitely lower triple-point temperature ($T_{\rm
tr}^*\simeq 0.0031$) and the as yet unpredicted reentrance of the
BCC phase when the FCC solid is isothermically expanded for reduced
temperatures in the $0.0031-0.0037$ range. In order to check whether
this latter feature is a spurious effect due to the finite size of
the system, we investigated the BCC-FCC phase coexistence also for
larger samples, but we did not register any significant change in
the location of the transition points (see the inset of
Fig.\,\ref{fig2}). In the triple-point region, the density jump is
$\sim 0.002$ across the fluid-solid transition and $\sim 0.00015$
across the solid-solid transition; the corresponding (absolute)
values of the entropy jump per particle are $\sim 0.7\,k_{\rm B}$
and $\sim 0.1\,k_{\rm B}$, respectively. The freezing line attains
its maximum value ($T_{\rm max}^*\simeq 0.00874$) for $\rho_{\rm
max}^*\simeq 0.239$. At the maximum, the fluid-solid transition is
still first-order with an entropy gap between the two phases equal
to $0.79\,k_{\rm B}$.

%
%
\begin{figure}
\includegraphics[width=8cm,angle=0]{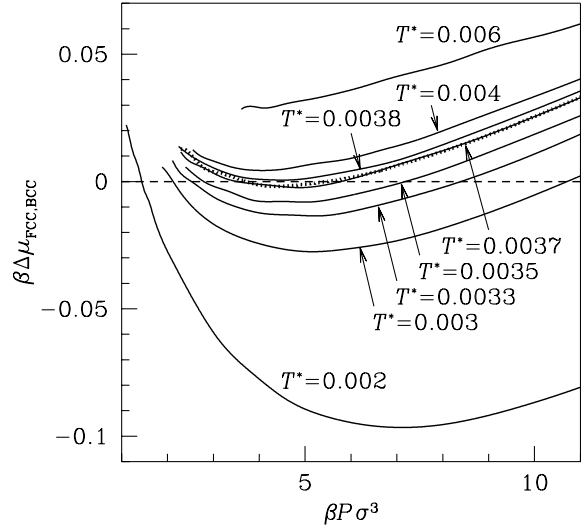}
\caption{\label{fig3} Temperature evolution of $\beta\Delta\mu_{\rm
FCC,\,BCC}$, the chemical-potential gap between the FCC ($N=1372$)
and BCC ($N=1458$) phases, plotted as a function of the pressure.
For $T^*=0.003$ and $0.006$, the systems investigated were smaller
($864$ and $1024$ particles, respectively). The dotted line
refers to a calculation carried out at $T^*=0.0037$ using larger
samples (FCC: $N=2048$; BCC: $N=2000$), and was plotted for a
comparison with the smaller-sizes calculation at the same
temperature.}
\end{figure}

To have a clue on the extent to which statistical errors may affect
our conclusions, we turn the reader's attention back to
Fig.\,\ref{fig1}. Let $\Delta\mu_{\rm A,\,B}(P)$ be the $\mu$ gap
(at fixed temperature) between two generic phases A and B. In the
triple-point region, we estimated a maximum statistical error on the
minimum value of $\beta\Delta\mu_{\rm FCC,\,BCC}(P)$ approximately
equal to $10^{-2}$. Of the same order is the maximum error we
estimated, near coexistence, on the values of $\beta\Delta\mu_{\rm
fluid,\,solid}(P)$. However, the rate of change of this latter
quantity is much larger, implying that its zero is more sharply
defined. This means that fluid-solid coexistence is numerically
better defined than solid-solid coexistence. If we follow the
evolution of $\beta\Delta\mu_{\rm FCC,\,BCC}(P)$ as a function of
$T$ (see Fig.\,\ref{fig3}), we realize that, over the whole
stability region of the reentrant BCC phase, this quantity takes
values that are of the order of the estimated numerical errors.
However, we can safely argue that the true errors are in fact much
smaller since, otherwise, we would have hardly obtained the very
smooth behavior represented in Fig.\,\ref{fig3} as well as the clear
phase portrait shown in the inset of Fig.\,\ref{fig2}. The already
mentioned absence of any significant size-dependence of
$\beta\Delta\mu_{\rm FCC,\,BCC}(P)$ at $T^*=0.0037$ is a further
guarantee of the reliability of the present phase diagram.

The FCC phase of the GCM is energetically favored at low densities
($\rho^*$ less than $\approx 0.17$), for temperatures up to
$T^*=0.008$. This may actually explain why the FCC-BCC coexistence
locus found in Ref.\,\cite{Lang} is an almost vertical line in the
$\rho$-$T$ plane, which leads to a more extended region of FCC
stability. In fact, Lang and coworkers used the Gibbs-Bogoliubov
inequality to optimize a strictly harmonic model of both solid
phases. This method may actually enhance the crystallinity and give,
at the same time, an inadequate representation of the entropic
contribution to the solid free energies. Considering that the
FCC-BCC transition occurs for rather small densities, the harmonic
approximation seems a severe limitation of the theory. In fact, the
variational technique basically propagates to higher temperatures
the relative stability condition valid at $T=0$. We also note that
the FCC-to-BCC transition undergone, with increasing temperatures,
by the GCM at low densities is to be ascribed to the higher entropy
of the BCC phase, that is likely due to the presence of a larger
number of soft shear modes. Even below the triple-point
temperature, BCC-ordered grains tend to form in the liquid that is
about to freeze, a phenomenon that substantially slows down
crystallization.

The low-density/low-temperature phase behavior of the GCM, with a
triple point separating a region where the fluid freezes into a FCC
structure from another region where these two phases are bridged by
an intermediate BCC phase, is rather common among model systems with
softly repulsive interactions, such as the inverse-power potential,
$v_n(r)=A(\sigma/r)^n$,~\cite{Laird,Agrawal} and the Yukawa
potential,
$v_\ell(r)=B\exp(-r/\ell)/r$.~\cite{Robbins,Meijer,Hamaguchi} The
phase diagram of the above two models is typically unfolded by one
or two (possibly rescaled) thermodynamic quantities and by the
relevant control parameter of the interaction, {\em i.e.}, the
inverse-power-law exponent $1/n$ or the Yukawa length $\ell$.
The crystalline pattern produced by such
potentials is critically determined by their degree of softness, the
BCC phase being promoted by a sufficiently soft interaction. A
criterion to relate the phase behavior of $v_n(r)$ and $v_\ell(r)$
to that of the GCM is to require the logarithmic derivatives of such
potentials to match that of the Gaussian potential (Eq.\,\ref{eq1}),
at least for separations close to the mean interparticle distance,
$\overline{r}=\rho^{-1/3}$. The values of $n$ and $\ell$ which
enforce this mapping are \be
\tilde{n}=2\rho^{*\,-2/3}\,\,\,\,\,\,{\rm and}\,\,\,\,\,\,
\tilde{\ell}=\frac{\rho^{*\,1/3}\sigma}{2-\rho^{*\,2/3}}\,.
\label{eq2} \ee When $\rho^*$ increases along the GCM freezing line,
the BCC phase becomes eventually stable. The same effect occurs with
the inverse-power and Yukawa potentials when $1/n$ and $\ell$ change
with the freezing density according to Eq.\,\ref{eq2}. In this
respect, $1/n$ and $\ell$ play a role analogous to that of an
{\it effective} temperature.
For instance, dilute solutions of charged colloids with
counterions~\cite{Sirota}, which constitute a practical realization
of $v_\ell(r)$, show this kind of behavior as a function of the
Debye screening length.

The phase diagram of the GCM shows some resemblance also with the
phase behavior of star-polymer solutions~\cite{Watzlawek}. At low
densities, the arm number $f$ plays the role of an effective inverse
temperature (for two reasons: the strength of the potential increases
with $f$ {\em and} its range increases with $f^{-1}$). For not too
huge values of $f$, the Yukawa repulsion yields a phase-stability
scenario that is very similar to that of the GCM.
Only for star-polymer packing fractions larger than $\sim 0.7$
(which is where the nearest-neighbor distance in a BCC solid is
$\sim\sigma$) will the peculiarities of the short-distance repulsion
make a difference, stabilizing other cubic phases that are likely
unstable in the GCM.

In this paper, we discussed the phase diagram of the GCM that was
redrawn using current best-quality numerical-simulation tools. We
predicted the existence of a narrow range of temperatures within
which the sequence of stable phases exhibited by the GCM upon
isothermal compression is fluid, BCC, FCC, BCC again, and finally
fluid again. We also rationalized these findings in terms of the
properties of other softly repulsive potentials, with an emphasis on
the phase behavior of star-polymer solutions.

We acknowledge some useful discussions with Gianpiero Malescio.

%
%

\end{document}